\documentclass{llncs}

\usepackage{algorithm} % ysmoon
\usepackage{algorithmic} % ysmoon
\usepackage{amssymb}
\usepackage{amsmath}
\usepackage{graphicx}

\usepackage{url}
\urldef{\mailsa}\path|{gils, ysmoon, bskim}@kangwon.ac.kr|%
\newcommand{\keywords}[1]{\par\addvspace\baselineskip
\noindent\keywordname\enspace\ignorespaces#1}

\newcommand{\OL}{\overline}

\begin{document}

\mainmatter  % start of an individual contribution

\title{Linear Detrending Subsequence Matching \\ in Time-Series Databases}
%\title{Noise Averaging Effect on Privacy-Preserving Clustering of Time-Series Data\vspace*{-0.20cm}}

%\titlerunning{Publishing Time-Series Data Under Preservation of Privacy and Distance Orders}

%\author{Myeong-Seon Gil, Yang-Sae Moon, Bum-Soo Kim, and Sang-Pil Kim}
\author{Myeong-Seon Gil, Yang-Sae Moon, and Bum-Soo Kim}
%
%\authorrunning{Lecture Notes in Computer Science: Authors' Instructionsa}

\institute{Department of Computer Science, Kangwon National University, Korea.\\%
           \mailsa\\ }


\toctitle{Lecture Notes in Computer Science}
\tocauthor{Authors' Instructions}
\maketitle%
%\vspace*{-0.40cm}%
\begin{abstract}
Each time-series has its own linear trend, the directionality of a time-series, and removing the linear trend is crucial to get the more intuitive matching results. Supporting the linear detrending in subsequence matching is a challenging problem due to a huge number of possible subsequences. In this paper we define this problem the {\it linear detrending subsequence matching\/} and propose its efficient {\it index-based\/} solution. To this end, we first present a notion of {\it LD-windows}\,(LD means linear detrending), which is obtained as follows: we eliminate the linear trend from a subsequence rather than each window itself and obtain LD-windows by dividing the subsequence into windows. Using the LD-windows we then present a lower bounding theorem for the index-based matching solution and formally prove its correctness. Based on the lower bounding theorem, we next propose the index building and subsequence matching algorithms for linear detrending subsequence matching. We finally show the superiority of our index-based solution through extensive experiments.
\end{abstract}

%---------------------------------------------------------------------------
\section{Introduction}\label{sec:intro}
%---------------------------------------------------------------------------

Time-series data are of growing importance in data mining and data warehousing\,\cite{Han07,Lian09}. A time-series is a sequence of real numbers representing values at specific points in time. Typical examples include stock prices, music data, network traffic data, moving object trajectories, and biomedical data\,\cite{Bar10,Bu09,Fal94,Keo06,Yi98}. The time-series data stored in a database are called {\it data sequences}, and those given by users are called {\it query sequences}. Finding data sequences similar to the given query sequence from the database is called {\it similar sequence matching\/} or {\it time-series matching}\,\cite{Han07}. In many similar sequence matching models, two sequences $X = \{X[1], \ldots, X[n[\}$ and $Y = \{Y[1], \ldots, Y[n]\}$ are said to be {\it similar\/} if the distance $D(X,Y) \leq \epsilon$, where $\epsilon$ is the user-specified tolerance. In this paper we use the Euclidean distance\,($= \sqrt{\sum_{i=1}^{n}|X[i]-Y[i]|^2}$) as the distance function of $D(X,Y)$.

\begin{figure}[hbt]
\centering
%  \vspace*{-0.30cm}%
  \includegraphics[width=12.0cm]{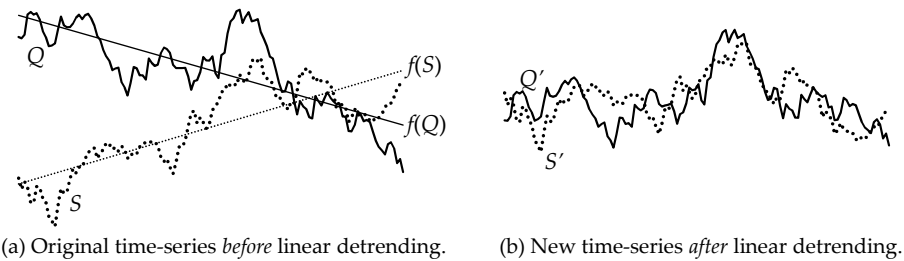}\\
%  \vspace*{-0.20cm}%
  \caption{Comparison of two sequences $S$ and $Q$ {\it before\/} and {\it after\/} linear detrending.}
%  \vspace*{-0.80cm}%
\label{fig:ldex}
\end{figure}

{\it Linear trend}, a representative distortion of time-series data\,\cite{Keo06,Loh10}, shows the directionality of a time-series, and {\it linear detrending} in similar sequence matching is crucial to get the more intuitive matching results. Figure \ref{fig:ldex} shows an example of comparing two sequences before and after linear detrending: Figure \ref{fig:ldex}(a) represents the original sequences $Q$ and $S$; Figure \ref{fig:ldex}(b) the linear detrended sequences $Q'$ and $S'$. We obtain $Q'$ and $S'$ by linear detrending, i.e., by subtracting the corresponding trend lines $f(Q)$ and $f(S)$ from $Q$ and $S$, respectively. In Figure \ref{fig:ldex}, there is a big distance between $Q$ and $S$, and these two sequences will be determined to be non-similar. In contrast, the distance between $Q'$ and $S'$ is very small in Figure \ref{fig:ldex}(b), and they will be determined to be similar. It means that non-similar sequences can be identified as similar ones after linear detrending, and vice versa. Likewise, linear detrending is useful to know similarity of changes which is hidden by the linear trend of time-series data\,\cite{Hill07,Keo06}. Motivated by this example, we attack the problem of linear detrending in similar sequence matching, especially in subsequence matching\,\cite{Fal94,Moon02}.

In this paper we address the problem of linear detrending in subsequence matching. Supporting the linear detrending is simple in whole matching since all data and query sequences have the same length. But, it is a challenging problem in subsequence matching because we need to consider a huge number of possible data subsequences to be linear detrended. We call this matching scheme the {\it linear detrending subsequence matching}. Formally speaking, for a query sequence $Q$ and a data sequence $S$, linear detrending subsequence matching finds all subsequences $S[i:j]$ such that $D(\OL{Q},\OL{S[i:j]}) \leq \epsilon$, where $\OL{Q}$ and $\OL{S[i:j]}$ are the linear detrended (sub)sequences of $Q$ and $S[i:j]$, respectively.

 We propose an {\it index-based\/} solution for linear detrending subsequence matching. To this end, we first present a novel notion of {\it LD-windows}, linear detrending-windows. Suppose a subsequence $S[i:j]$ include a window $S[a:b]$ (i.e., $i \leq a < b \leq j$), then we obtain the LD-window of $S[a:b]$ by eliminating the linearity of  subsequence $S[i:j]$ rather than that of window $S[a:b]$ itself. This notion enables an LD-window to represent multiple subsequences of different lengths, and eventually, we can use only one index in subsequence matching\,\cite{Moon07b}. Using the LD-windows we next present a lower bounding theorem for the index-based matching solution and formally prove its correctness. Based on this lower bounding theorem, we then propose the index building and subsequence matching algorithms, respectively. We finally showcase the superiority of our index-based solution through extensive experiments. Experimental results show that, compared with the sequential scan, our solution improves the matching performance by one or two orders of magnitude.

%---------------------------------------------------------------------------
\section{Related Work} \label{sec:related}
%---------------------------------------------------------------------------
%
Similar sequence matching can be classified into whole matching and subsequence matching\,\cite{Moon02}. The whole matching\,\cite{Agr93,Chan03} finds data sequences similar to a query sequence, where the lengths of data and query sequences are all identical. The subsequence matching\,\cite{Fal94,Han07,Loh04,Loh10,Moon07a,Moon07b} finds subsequences, contained in data sequences, similar to a query sequence of arbitrary length. Likewise, subsequence matching is a generalization of whole matching\,\cite{Fal94,Moon02}, and we thus focus on subsequence matching in this paper.

Many efficient solutions have been proposed for subsequence matching\,\cite{Fal94,Han07,Moon02}. These solutions consists of index-building and subsequence matching algorithms. In the index-building algorithm, the solution constructs an $R^*$-tree as follows: it divides data sequences into multiple windows of size $\omega$; transforms those windows to $f(\ll \omega)$-dimensional points using the lower-dimensional transformation such as discrete Fourier transform\,(DFT) and piecewise aggregate approximation\,(PAA); and stores the points (or minimum bounding rectangls\,(MBRs) containing multiple points) into the R$^*$-tree. In the subsequence matching algorithm, the solution finds similar subsequences as follows: it divides the query sequence into multiple windows of the same size $\omega$; transforms each window to an $f$-dimensional point; makes a range query using the point and the tolerance; constructs a candidate set by searching the $R^*$-tree; and finally obtain actual similar subsequences by eliminating {\it false alarms\/} through the post-processing step\,\cite{Agr93,Fal94,Loh10}.

Representative distortions embedded in time-series are offset translation, amplitude scaling, noise, and linear trend\,\cite{Fu08,Keo06}. In similar sequence matching, there have been many efforts to remove these distortions from time-series data. For example, offset translation and amplitude scaling can be solved by the normalization transform, and its subsequence matching solutions were proposed in \cite{Loh04,Moon07b,Loh10}. Also, the moving average transform can alleviate noise of time-series, and its subsequence matching solution was proposed in \cite{Moon07a}. To our best knowledge, however, there is no solution to linear detrending subsequence matching, and in this paper we define the problem first and present an efficient index-based solution.

%---------------------------------------------------------------------------
\section{Linear Detrending Subsequence Matching} \label{sec:ldsm}
%---------------------------------------------------------------------------
%
\subsection{Problem Definition}
For a time-series, its linear trend is a straight line that most likely reflects its directionality. The {\it least square method\/} is most widely used to obtain the line of a time-series\,\cite{Hill07}. For a sequence $X = \{X[1], \ldots, X[n]\}$, a linear function by the least square method is given by $g(k) = \alpha k + \beta$, where $\alpha$ and $\beta$ are obtained by Eq. (\ref{eq:lsm})\,\cite{Hill07}.
\begin{eqnarray}
    \alpha & = & \frac{n\sum_{k=1}^{n}kX[k]-\sum_{k=1}^{n}k\cdot\sum_{k=1}^{n}X[k]}{n\sum_{k=1}^{n}k^2-(\sum_{k=1}^{n}k)^2}, \nonumber \\%
    \beta & = & \frac{\sum_{k=1}^{n}X[k]}{n} - \alpha\frac{\sum_{k=1}^{n}k}{n}. \label{eq:lsm}
\end{eqnarray}
{\it Linear detrending\/} is the process of obtaining a new time-series from an original time-series by removing the corresponding linear trend. The following is the formal definition of linear detrending.
\begin{definition}
\label{def:ld}
For a sequence $X = \{X[1], \ldots X[n]\}$ and its trend line $g(k) = \alpha k + \beta$, the {\it linear detrending sequence\/} of $X$, {\it LD-sequence\/} of $X$ in short, is defined as $\OL{X} = \{\OL{X}[1], \ldots, \OL{X}[n]\}$, where $\OL{X}[k] = X[k] - g(k), k = 1, 2, \ldots, n$. $\Box$
\end{definition}

Linear detrending is simply solved in whole matching, but it is a challenging problem in subsequence matching. In whole matching, the lengths of data and query sequences are all identical, and we simply use the whole matching solution\,\cite{Agr93} after linear detrending of all time-series. In contrast, the solution is not simple in subsequence matching by the following reasons: (1) data subsequences in different positions have different linear trend even though they have the same length; and (2) data subsequences of different lengths also have different linear trend even though they start from the same position. Therefore, we need to consider different linear trend for all possible query lengths and for all possible positions, and we cannot use the traditional whole/subsequence matching solutions for linear detrending subsequence matching.

We formally define the problem of linear detrending subsequence matching. We first present similarity of two sequences by considering the linear detrending.
\begin{definition}
\label{def:ld-sim}
For two sequences $X$ and $Y$ of the same length and their LD-sequences $\OL{X}$ and $\OL{Y}$, we define that $X$ and $Y$ (or $\OL{X}$ and $\OL{Y}$) are {\it LD-similar\/} if the Euclidean distance between $\OL{X}$ and $\OL{Y}$ is less than or equal to the tolerance $\epsilon$, i.e., if $D(\OL{X},\OL{Y}) \leq \epsilon$. $\Box$
\end{definition}
Using the concept of LD-similarity, we now define the problem of linear detrending subsequence matching as follows:
\begin{definition}
\label{def:ldsm}
For a data sequence $S$, a query sequence $Q$, and the tolerance $\epsilon$, {\it linear detrending subsequence matching\/} is the problem of finding all subsequences $S[i:j]$ which are LD-similar to $Q$, i.e., finding all subsequences $S[i:j]$ such that $D(\OL{Q},\OL{S[i:j]}) \leq \epsilon$. $\Box$
\end{definition}

\subsection{Sequential Scan-based Solution and Its Problems}
Sequential scan accesses every subsequence $S[i:j]$ sequentially and investigates its LD-similarity by computing $D(\OL{Q},\OL{S[i:j]})$. Algorithm \ref{alg:seqscan} shows the sequential scan algorithm, {\sf LDSeqScan}, which is simple and self-explained. Algorithm {\sf LDSeqScan} accesses all possible subsequences one by one in Lines 3 to 7 and returns LD-similar subsequences by investigating the LD-similarity.

\begin{algorithm}[hbt]
\caption{{\sf LDSeqScan}(data sequence $S$, query sequence $Q$, tolerance $\epsilon$)}\label{alg:seqscan}%
%\begin{scriptsize}
\begin{algorithmic}[1]
\STATE Compute a trend line $g(k)$ from $Q$ using the least square method; %
\STATE Obtain $\OL{Q}$ from $Q$ and $g(k)$ through linear detrending;%
\STATE {\bf for each} subsequence $S[i:j]$ of length $Len(Q)$ {\bf do};%
\STATE ~~~~Compute a trend line $g'(k)$ from $S[i:j]$ using the least square method;
\STATE ~~~~Obtain $\OL{S[i:j]}$ from $S[i:j]$ and $g'(k)$ through linear detrending;
\STATE ~~~~Return the subsequence $S[i:j]$ if $D(\OL{Q},\OL{S[i:j]}) \leq \epsilon$; ~~// LD-similar
\STATE {\bf end-for}
\end{algorithmic}
%\end{scriptsize}
\end{algorithm}

The sequential scan algorithm has an advantage of simplicity, but has a disadvantage of incurring severe CPU and I/O overhead. First, the algorithm causes many disk accesses since it accesses an entire data sequence in a database. Second, the algorithm also causes severe CPU overhead since it investigates the LD-similarity for every individual subsequence by performing the linear detrending and by computing the Euclidean distance. This CPU and I/O overhead makes {\sf LDSeqScan} impractical for a large time-series database. To solve this problem, we propose an efficient index-based solution in the next Section \ref{subsec:ibsol}.

\subsection{Index-based Solution and Its Algorithms} \label{subsec:ibsol}
As in the traditional subsequence matching\,\cite{Fal94,Han07,Moon02}, we use the window construction mechanism that divides data and query sequences into disjoint/sliding windows of the fixed size. However, our solution quite differs from the traditional ones in constructing windows due to use of linear detrending. Each window should be mapped to multiple windows in the linear detrending subsequence matching while it does not in the traditional subsequence matching. This is because, in linear detrending subsequence matching, each window has multiple trend lines by different lengths and different positions of subsequences that include the window. Formally speaking, for a given window $S[a:b]$, there are many different subsequences $S[i:j]$'s that include $S[a:b]$; their trend lines are also different from each other; and the window $S[a:b]$ is mapped to multiple windows due to different trend lines. We call this complex property the {\it multiple mapping property}, which was already presented in the normalization-transformed subsequence matching\,\cite{Moon07b}. The traditional subsequence matching solutions\,\cite{Fal94,Han07,Moon02} do not have the multiple mapping property, but we need to support this property in linear detrending subsequence matching.

To support the multiple mapping property, for a given window, we do not remove the linear trend of the window itself, but we instead remove the linear trend of a subsequence including that window. To this end, we present a notion of LD-windows as follows:
\begin{definition}
\label{def:ld-win}
Suppose $S[i:j]$ be a subsequence of a sequence $S$, $g(k)$ be a linear function of $S[i:j]$, and $S[a:b]$ be a window included in $S[i:j]$, then {\it LD-windows\/} of $S[a:b]$ against $S[i:j]$, denoted by $\OL{S_{\{i,j\}}[a:b]}$, is defined as a new window whose entry $\OL{S_{\{i,j\}}[k]}(k = a, a+1, \ldots, b)$ is set to $S[k] - g(k)$. $\Box$
\end{definition}
Definition \ref{def:ld-win} means that a window $S[a:b]$ is mapped to an LD-window $\OL{S_{\{i,j\}}[a:b]}$ by the trend line of a subsequence $S[i:j]$ which includes $S[a:b]$. Because of the multiple mapping property, there are many subsequences $S[i:j]$'s that include $S[a:b]$, and thus, each window $S[a:b]$ is mapped to multiple LD-windows $\OL{S_{\{i,j\}}[a:b]}$'s for different subsequences $S[i:j]$'s.

\begin{figure}[hbt]
\centering
%  \vspace*{-0.30cm}%
  \includegraphics[width=12.0cm]{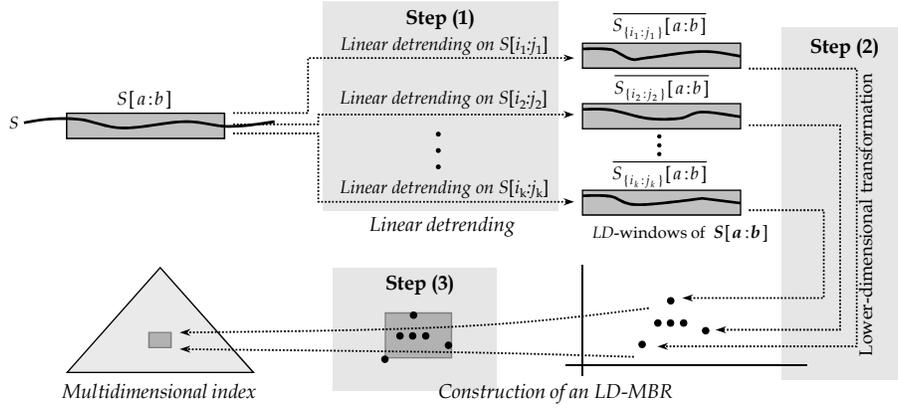}\\
%  \vspace*{-0.20cm}%
  \caption{Process of constructing an LD-MBR of a window $S[a:b]$.}
%  \vspace*{-0.80cm}%
\label{fig:ld-mbr}
\end{figure}

Like the traditional subsequence matching algorithms\,\cite{Fal94,Han07,Moon02}, our index-based solution first transforms each {\it high-dimensional\/} window to a {\it low-dimensional\/} point and then stores the point into the multidimensional index. Unlike the traditional algorithms, however, our solution maps each high-dimensional window to a low-dimensional MBR that bounds multiple low-dimensional points. This is due to the multiple mapping property that a window is mapped to multiple LD-windows. Constructing an MBR from a window is performed as follows: (1) the given window is mapped to multiple LD-windows; (2) LD-windows are transformed to low-dimensional points by the lower-dimensional transformation; and (3) a low-dimensional MBR is constructed by bounding the transformed points. We call this MBR {\it LD-MBR\/} and formally define it as follows:
\begin{definition}
\label{def:ld-mbr}
Suppose $s$ be a window of a sequence $S$, $\mathbb{S}$ be \{$\OL{s}$ $|$ $\OL{s}$ is an LD window of $s$\}, and $T(\cdot)$ be a function of lower-dimensional transformation, then {\it LD-MBR\/} of $s$, denoted by  $\mathbb{M}(T(\mathbb{S})$), is defined as a low-dimensional MBR that bounds all low-dimensional points $T(\OL{s})$ for all $\OL{s} \in \mathbb{S}$. $\Box$
\end{definition}
Figure \ref{fig:ld-mbr} shows the process of constructing an LD-MBR of a window $S[a:b]$. The process is as follows: (1) the window $S[a:b]$ is mapped to multiple LD-windows $\OL{S_{\{i_k:j_k\}}[a:b]}$'s by considering possible subsequences $S[i_k:j_k]$'s; (2) each LD-window is transformed to a low-dimensional point; and (3) an LD-MBR is constructed by bounding those points.

Our index-based solution is developed from the following Theorem \ref{th:ldsm}.
\begin{theorem}
\label{th:ldsm}
{\it For a query sequence $Q$, a data subsequence $S[i:j]$, a tolerance $\epsilon$, a function $T(\cdot)$ of lower-dimensional transformation, if $Q$ and $S[i:j]$ are LD-similar, that is, if $D(\OL{Q},\OL{S[i:j]}) \leq \epsilon$, the distance between $T(\OL{q_k})$ and $\mathbb{M}(T(\mathbb{S}_k))$ $\leq \epsilon/\sqrt{p}$, where $\OL{s_1}, \ldots, \OL{s_p}$ and $\OL{q_1}, \ldots, \OL{q_p}$ are $p$ disjoint windows of $\OL{Q}$ and $\OL{S[i:j]}$, respectively, and $\mathbb{S}_k$ is the set of LD-windows of $s_k$. That is, the following Eq. {\rm (\ref{eq:ldsm})} holds:}
\begin{equation}
\label{eq:ldsm}
D(\OL{Q},\OL{S[i:j]})
  \leq \epsilon \Longrightarrow \bigvee_{k=1}^{p}D(T(\OL{q_k}),\mathbb{M}(T(\mathbb{S}_k))) \leq \epsilon/\sqrt{p}.
\end{equation}
{\sc Proof}: The proof is similar to that of {\it normalization-transformed subsequence matching\/} in the previous work\cite{Moon07b}. Refer to the proof of Theorem 1 in \cite{Moon07b} for the detailed proof. $\Box$
\end{theorem}
Theorem \ref{th:ldsm} guarantees correctness of our index-based solution to linear detrending subsequence matching. Like the traditional subsequence matching solutions, our solution also consists of two algorithms: (1) the index-building algorithm and (2) the subsequence matching algorithm.

Algorithm \ref{alg:bi} shows the index-building algorithm. In Line 2 we divide the given data sequence into sliding or disjoint windows of size $\omega$. For the first subsequence matching solution of \cite{Fal94}, we use the sliding window; in contrast, for the recent Dual\,Match\,\cite{Moon02}, we use the disjoint window. In Lines 4 to 14, we build a multidimensional index by repeating the following three steps for each window $S[a:b]$: (1) compute trend lines of all possible subsequences\,(Line 8); obtain LD-windows using those trend lines\,(Line 9); and (3) map those LD-windows to an LD-MBR\,(Line 10). After obtaining an LD-MBR from a window, we store it into the index with its starting offset $a$\,(Line 13). Once we build an index by Algorithm BuildIndex, we use it repeatedly in the subsequence matching algorithm.

Algorithm \ref{alg:sm} shows the subsequence matching algorithm. In Line 2 we first eliminate the linear trend from the query sequence $Q$. In Line 3 we divide the LD sequence $\OL{Q}$ into disjoint or sliding windows $\OL{q}$ of size $\omega$. For the first solution of \cite{Fal94}, we use the disjoint window; in contrast, for Dual\,Match\,\cite{Moon02}, we use the sliding window. In Lines 5 to 11, we find candidate subsequences by repeating the following steps for each query window $\OL{q}$: (1) transform a high-dimensional window $\OL{q}$ to a low-dimensional point\,(Line 6); (2) make a range query using that point and the given tolerance\,(Line 7); and (3) find candidate subsequences by evaluating the range query on the index\,(Lines 8 and 9). After obtaining the candidate subsequences, we finally perform the post-processing step\,\cite{Agr93,Fal94,Han07,Moon02} to identify  true LD-similar subsequences by accessing actual subsequences and eliminating false alarms.

\begin{algorithm}[hbt]
\caption{{\bf BuildIndex}(data sequence $S$)}\label{alg:bi}%
%\begin{scriptsize}
\begin{algorithmic}[1]
\STATE Let the window size be $\omega$ and the maximum/minimum query lengths be $l_{\it min}$, $l_{\it max}$;%
\STATE Divide $S$ into windows of size $\omega$;%
\STATE ~~~~// {\it sliding\/} windows for \cite{Fal94}; {\it disjoint\/} windows for Dual\,Match\,\cite{Moon02}.%
\STATE {\bf for each} window $S[a:b]$ in $S$ {\bf do}%
\STATE ~~~~Make an $f$-dimensional MBR $\mathbb{M}$ which is initially empty;
\STATE ~~~~{\bf for each} query length $l \in [l_{\it min},l_{\it max}]$ {\bf do}
\STATE ~~~~~~~~{\bf for each} subsequence $S[i:j]$ of length $l$ that includes $S[a:b]$ {\bf do}
\STATE ~~~~~~~~~~~~Compute a trend line of $S[i:j]$ based on the least square method;
\STATE ~~~~~~~~~~~~Obtain the LD-window $\OL{S_{\{i,j\}}[a:b]}$; // linear detrending
\STATE ~~~~~~~~~~~~Transform $\OL{S_{\{i,j\}}[a:b]}$ to an $f$-dimensional point and include it into $\mathbb{M}$;
\STATE ~~~~~~~~{\bf end-for}
\STATE ~~~~{\bf end-for}
\STATE ~~~~Make a record $<$$\mathbb{M}$, offset = $a$$>$ for $S[a:b]$, and store it into the index;
\STATE {\bf end-for}
\end{algorithmic}
%\end{scriptsize}
\end{algorithm}

\begin{algorithm}[hbt]
\caption{{\bf SubsequenceMatching}(query sequence $Q$, tolerance $\omega$)}\label{alg:sm}%
%\begin{scriptsize}
\begin{algorithmic}[1]
\STATE Let the window size be $\omega$;  // $\omega$ is the same one used in Algorithm \ref{alg:bi}.
\STATE Obtain $\OL{Q}$ from $Q$ by eliminating the linear trend;
\STATE Divide $\OL{Q}$ into windows of size $\omega$;
\STATE ~~~~// {\it disjoint\/} windows for \cite{Fal94}; {\it sliding\/} windows for Dual\,Match\,\cite{Moon02}.%
\STATE {\bf for each} window $\OL{q}$ {\bf do}
\STATE ~~~~Transform $\OL{q}$ to an $f$-dimensional point; // lower-dimensional transformation
\STATE ~~~~Construct a range query using that point and $\epsilon/\sqrt{p}$;
\STATE ~~~~~~~~// $p$ is the number of included windows in $Q$\,\cite{Moon02}.
\STATE ~~~~Evaluate the query on the index and find the record of the form $<$$\mathbb{M},a$$>$;
\STATE ~~~~Include in the candidate set the subsequences $S[i:j]$ obtained from $<$$\mathbb{M},a$$>$;
\STATE {\bf end-for}
\STATE Perform the post-processing step\,\cite{Agr93,Fal94,Han07,Moon02} to eliminate false alarms;
\end{algorithmic}
%\end{scriptsize}
\end{algorithm}

%---------------------------------------------------------------------------
\section{Experimental Evaluation} \label{sec:exp}
%---------------------------------------------------------------------------
%

%
\subsection{Experimental Setup}
We have performed experiments using three real data sets, which also used in the previous work\,\cite{Keo06}. A data set consists of a long data sequence and has the same effect as the one consisting of multiple data sequences\,\cite{Fal94,Moon02}. The first data set contains electrocadiogram\,(ECG) data, and we call this data set {\it ECG-DATA}. The second data set shows tax growth rates, and we call this data set {\it TAX-DATA}. The third data set contains exchange rate data, and we call this data set {\it EXCH-DATA}. The length of each data set is 100,000, that is, each data set consists of 100,000 entries\,(time points).

In the experiment we have compared two matching solutions: (1) {\sf LDSeqScan}, a sequential scan solution presented in Section 3.2; (2) an index-based matching solution proposed in Section 3.3. We have adopted the first subsequence matching solution of \cite{Fal94} and implemented our index-based approach to that subsequence matching solution. For simplicity, we call this index-based solution {\sf LDIndexMatch}. We have performed two different experiments. In the first experiment we set the window size and the selectivity\footnote{Selectivity = $\frac{\text{the number of subsequences that are LD-similar with the query sequence}}{\text{the number of all possible subsequences in the database}}$}\,\cite{Fal94,Moon02} to 256 and $10^{-3}$, respectively, and vary the query sequence length from 256 to 1024. In the second experiment we set the window size and the query sequence length to 256 and 512, respectively, and vary the selectivity from $10^{-1}$ to $10^{-4}$. We obtain the desired selectivity by controlling the tolerance $\epsilon$\,\cite{Moon02}. As the metric of efficiency, we measure the elapsed time of each solution. We generate query sequences from the data sequence by taking subsequences of length ${\it Len}(Q)$ starting from random offsets\,\cite{Fal94,Moon02,Moon07b}. To avoid effects of noise, we experiments with 20 different query sequences of the same length and use their average as the result.

The hardware platform was SUN Ultra 25 workstation equipped with UltraSPARC IIIi 1.34GHz CPU, 1.0GB RAM, and an 80GB hard disk; its software platform was Solaris 10. We used C/C++ language for implementing two matching solutions. In {\sf LDIndexMatch}, we used PAA\,\cite{Han07,Keo06} as the lower-dimensional transformation and extracted eight features from an window of size 256. As the multidimensional index, we used the R$^*$-tree\,\cite{Agr93,Fal94,Moon02} for {\sf LDIndexMatch}.

\subsection{Experimental Results}
\begin{figure}[hbt]
\centering
  %\vspace*{-0.30cm}%
  \includegraphics[width=11.5cm]{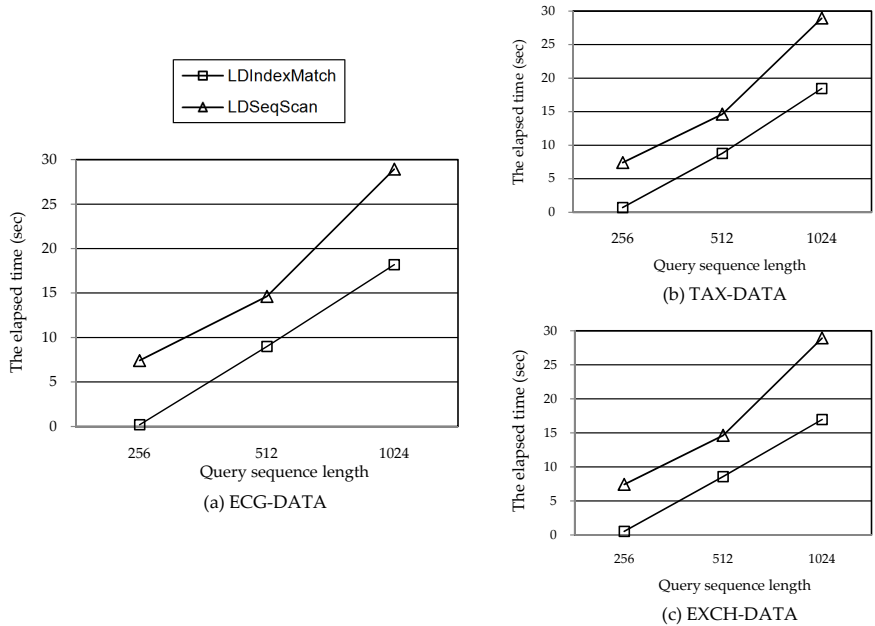}\\
  %\vspace*{-0.20cm}%
  \caption{Experimental results by varying the query sequence length.}
  %\vspace*{-0.40cm}%
\label{fig:query}
\end{figure}

Figure \ref{fig:query} shows the results of the first experiment that uses different lengths of query sequences. We first note that, in Figure \ref{fig:query}(a) of ECG-DATA, {\sf LDIndexMatch} significantly outperforms {\sf LDSeqScan}. This means that the notion of LD-windows works properly, and it prunes many unnecessary accesses on subsequences at the index level. As shown in Figure \ref{fig:query}(a), as the query sequence length decreases, the performance difference between two solutions becomes larger. For example, compared with {\sf LDSeqScan}, {\sf LDIndexMatch} reduces the elapsed time by 38.0 times for the query sequence of length 1024; in contrast, it reduces the elapsed time by 1.60 times only for the query sequence of length 256. This is explained by the {\it window size effect\/}\cite{Moon02} that the performance of index-based solutions decreases as the query sequence length on the given window size increases. We can solve this problem by using the {\it index interpolation\/} technique\,\cite{Loh04} which uses multiple indexes (for multiple window sizes) to obtain the better performance. Figures \ref{fig:query}(b) and \ref{fig:query}(c) of TAX-DATA and EXCH-DATA show the very similar trend with Figure \ref{fig:query}(a) of ECG-DATA. It means that the proposed {\sf LDIndexMatch} exploits the pruning effect efficiently, regardless of data types. In summary of Figure \ref{fig:query}, our index-based solution, {\sf LDIndexMatch}, improves the overall performance by 1.57 to 38.0 times compared with the straightforward solution, {\sf LDSeqScan}.

\begin{figure}[hbt]
\centering
  %\vspace*{-0.30cm}%
  \includegraphics[width=11.5cm]{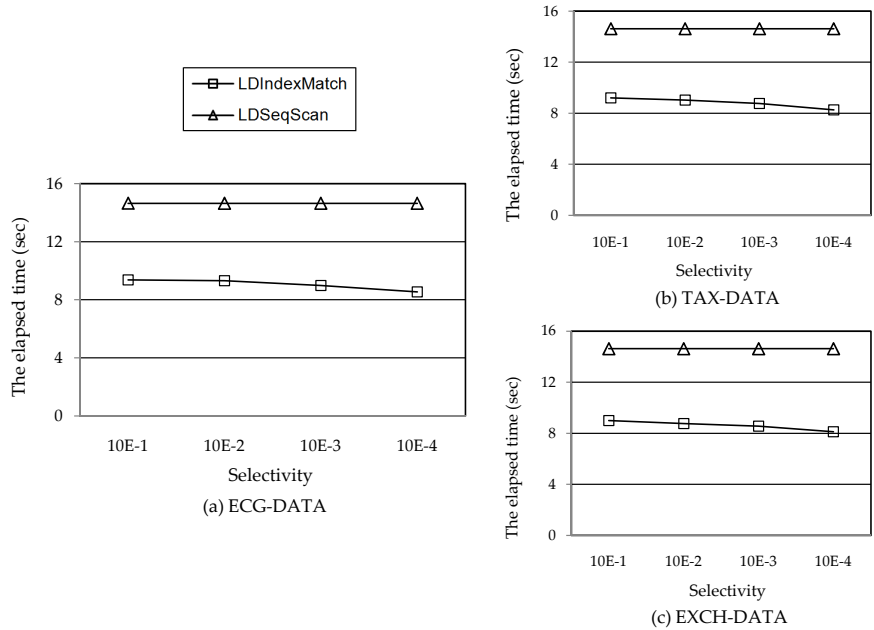}\\
  %\vspace*{-0.20cm}%
  \caption{Experimental results by varying the selectivity.}
  %\vspace*{-0.40cm}%
\label{fig:tol}
\end{figure}

Figure \ref{fig:tol} shows the results of the second experiment that uses difference selectivities (i.e., different tolerances). As in Figure \ref{fig:query}, {\sf LDIndexMatch} also outperforms {\sf LDSeqScan} in all selectivity ranges of Figure \ref{fig:tol}. We note that, in Figure \ref{fig:tol}(a) of ECG-DATA, the performance difference between {\sf LDIndexMatch} and {\sf LDSeqScan} increases as the selectivity decreases. This is because, as the selectivity decreases, the number of candidate subsequences also decreases. More precisely, as shown in Lines 7 to 10 of Algorithm 3, the smaller selectivity incurs the smaller number of candidate subsequences at the index level, and this reduces the false alarms that cause the disk accesses and the expensive computations on the actual subsequences. As in Figure \ref{fig:query}, Figures \ref{fig:tol}(b) and \ref{fig:tol}(c) of TAX-DATA and EXCH-DATA show the very similar trend with Figure \ref{fig:tol}(a) of ECG-DATA. Figure \ref{fig:tol} demonstrates that {\sf LDIndexMatch} beats {\sf LDSeqScan} in all selectivity ranges, and this means that {\sf LDIndexMatch} does not much depend on the selectivity values.

%---------------------------------------------------------------------------
\section{Conclusions} \label{sec:con}
%---------------------------------------------------------------------------
%
In this paper we introduced a new problem of linear detrending subsequence matching and proposed an efficient index-based solution. Contributions of the paper are summarized as follows. First, we formally defined the linear detrending subsequence matching and presented its sequential scan-based solution. Second, we presented a novel notion of {\it LD-windows}, and using LD-windows we proposed an index-based solution. We here formally proved correctness of our index-based solution. Third, we described the index-building and subsequence matching algorithms of the index-based solution. Fourth, we showcased that, compared with the straightforward sequential scan, our index-based solution significantly improved the matching performance by one or two orders of magnitude. We believe that the linear detrending subsequence matching and its index-based solution will be very helpful to find meaningful time-series patterns hidden by the linear trend.

%ACKNOWLEDGMENTS are optional
\section*{Acknowledgments}
This work was partially supported by Defense Acquisition Program Administration and Agency for Defense Development under the contract. (UD060048AD)

%---------------------------------------------------------------------------

\end{document}